\documentclass[journal,twoside,final]{IEEEtran}
\usepackage{amssymb}
\usepackage{graphicx}
\usepackage{url}

\graphicspath{{./}}	

\begin{document}

\title{\LARGE Comments on ``IEEE 1588 Clock Synchronization using Dual Slave
  Clocks in a Slave''}

\author{Kyeong Soo Kim, \IEEEmembership{Member, IEEE}\\
  \thanks{K. S. Kim is with the Department of Electrical and Electronic
    Engineering, Xi'an Jiaotong-Liverpool University, Suzhou 215123, Jiangsu
    Province, P. R. China (e-mail: Kyeongsoo.Kim@xjtlu.edu.cn).}
}

\maketitle


\begin{abstract}
  In the above letter, Chin and Chen proposed an IEEE 1588 clock synchronization
  method based on dual slave clocks, where they claim that multiple unknown
  parameters --- i.e., clock offset, clock skew, and master-to-slave delay ---
  can be estimated with only one-way time transfers using more equations than
  usual. This comment investigates Chin and Chen's dual clock scheme with
  detailed models for a master and dual slave clocks and shows that the
  formulation of multi-parameter estimation is invalid, which affirms that it is
  impossible to distinguish the effect of delay from that of clock offset at a
  slave even with dual slave clocks.
\end{abstract}


\section*{\(~\)}
\label{sec-1}
\IEEEPARstart{D}{ual} slave clocks method is proposed to overcome the limit of
conventional IEEE 1588 clock synchronization approaches causing clock
synchronization errors in asymmetric links by way of using only one-way exchange
of timing messages \cite{chin09:_ieee}, where the authors claim that multiple
unknown parameters in clock synchronization --- i.e., clock offset, clock skew,
and master-to-slave delay --- can be estimated simultaneously using more
equations than usual resulting from the use of dual slave clocks. The claim,
however, is contrary to the well-known fact that the clock offset and the delay
cannot be differentiated with only one-way message dissemination
\cite{wu11:_clock_synch_wirel_sensor_networ}.

To investigate the issues in the formulation of multi-parameter estimation in
\cite{chin09:_ieee}, we first model the master clock and the dual slave clocks
generated by a common signal in terms of an \emph{ideal, global reference time}
\(t\).
For simplicity, we consider continuous clock models and ignore clock jitters in
modeling; in this case, the master clock and the dual slave clocks at $t$ are
given by
\begingroup
\setlength{\arraycolsep}{0.0em}
\begin{eqnarray}
  T_{m}(t) &{=}& f_{m} \cdot t + \theta_{m}, \label{eq:ref_master_clock} \\
  T_{s1}(t) &{=}& 2f_{s} \cdot t + \theta_{s,1}, \label{eq:ref_slave_clock1} \\
  T_{s2}(t) &{=}& f_{s} \cdot t + \theta_{s,2}, \label{eq:ref_slave_clock2}
\end{eqnarray}
\endgroup
where $f_{m}$ and $f_{s}$ are the frequencies of the master clock and the common
clock driving the dual slave clocks, and $\theta_{m}$, $\theta_{s,1}$, and
$\theta_{s,2}$ are the phase differences of the master clock and the dual slave
clocks (i.e., clock 1 and clock 2 shown in Fig.\(~\)3
of \cite{chin09:_ieee}) with respect to the ideal reference clock, respectively.

Given these models, we can describe the two slave clocks in terms of the master
clock (i.e., $T_{m}$) as follows:
\begingroup
\setlength{\arraycolsep}{0.0em}
\begin{eqnarray}
  T_{s1}(T_{m}) &{=}& 2\left(1 + \epsilon\right) T_{m} + \tilde{\theta}_{s,1}, \label{eq:master_slave_clock1} \\
  T_{s2}(T_{m}) &{=}& \left(1 + \epsilon\right) T_{m} + \tilde{\theta}_{s,2}, \label{eq:master_slave_clock2}
\end{eqnarray}
\endgroup
where
\begingroup
\setlength{\arraycolsep}{0.0em}
\begin{eqnarray}
  \epsilon &{=}& \frac{f_{s}-f_{m}}{f_{m}}, \label{eq:clock_skew} \\
  \tilde{\theta}_{s,1} &{=}& \theta_{s,1} - 2\left(1 + \epsilon\right)\theta_{m}, \label{eq:slave_clock1_phase} \\
  \tilde{\theta}_{s,2} &{=}& \theta_{s,2} - \left(1 + \epsilon\right)\theta_{m}. \label{eq:slave_clock2_phase}
\end{eqnarray}
\endgroup
Note that at the slave, $\epsilon$ (i.e., normalized clock skew) and
$\theta_{m}$ are unknown because these are parameters related with the master
clock, while $\theta_{s,1}$ and $\theta_{s,2}$ --- though their true values are
also unknown --- are controllable and their ratio (i.e.,
$\theta_{s,1}/\theta_{s,2}$) can be set to the frequency ratio between the slave
clocks due to the dual clock generation described in \cite{chin09:_ieee} (i.e.,
2 when both clocks begin at the same time; see Fig.\(~\)\ref{fg:phase_ratio}
for illustration).
\newlength{\figwidth} \setlength{\figwidth}{.65\linewidth}
\begin{figure}[!t]
  \begin{center}
    \includegraphics[width=\the\figwidth]{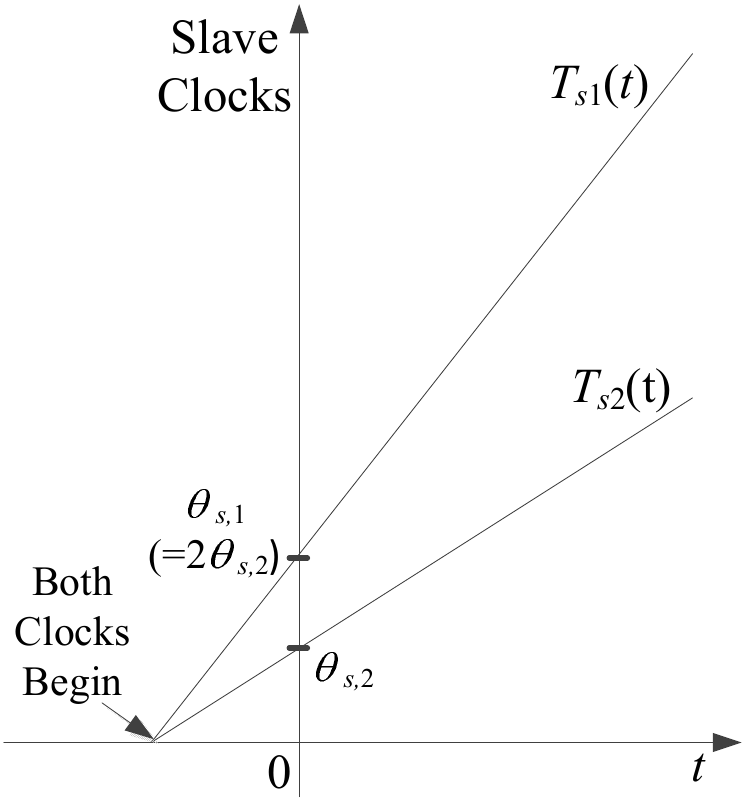}
  \end{center}
  \caption{Illustration of the ratio of phase differences (w. r. t. the ideal
    reference clock) when dual slave clocks begin simultaneously.}
  \label{fg:phase_ratio}
\end{figure}

Now consider the equations (5) and (6) of \cite{chin09:_ieee} describing a
relationship between the times of slave clocks when the $i$-th Sync message is
received:
\begingroup
\setlength{\arraycolsep}{0.0em}
\begin{eqnarray}
  T_{s1}.i &{=}& 2 \left(1 + \epsilon\right) \left(T_{m}.i + D_{m2s}\right) + \theta + \phi_{1,i} \label{eq:slave_clock1} \nonumber \\
  T_{s2}.i &{=}& \left( 1 + \epsilon\right) \left(T_{m}.i + D_{m2s}\right) + \theta + \phi_{2,i}  \label{eq:slave_clock2} \nonumber ,
\end{eqnarray}
\endgroup
where $D_{m2s}$ is the master-to-slave delay, $\theta$ is a common offset of the
slave clocks, and $\phi_{1,i}$ and $\phi_{2,i}$ are random jitters of slave
clock 1 and 2 within a period, respectively. We can see that, except for the
noise components (i.e., $\phi_{1,i}$ and $\phi_{2,i}$ representing clock
jitters), (\ref{eq:master_slave_clock1}) and (\ref{eq:master_slave_clock2}) are
generalized expressions (i.e., at any time instant) for time relationship
between the master and the slave clocks.

Chin and Chen argue that, because a single signal drives both the clocks as
shown in Fig. 3 of \cite{chin09:_ieee}, they have the same offset (i.e.,
$\theta$) with respect to the master clock. This, however, is not the case in
general; from (\ref{eq:slave_clock1_phase}) and (\ref{eq:slave_clock2_phase}),
we obtain the condition for the slave clocks to have a common offset with
respect to the master clock (i.e., $\tilde{\theta}_{s,1}=\tilde{\theta}_{s,2}$)
as follows:
\begin{equation}
  \label{eq:phase_condition}
  \theta_{s,1} - \theta_{s,2} = \left(1 + \epsilon\right)\theta_{m} .
\end{equation}
As we discussed with (\ref{eq:clock_skew})--(\ref{eq:slave_clock2_phase}), the
number of parameters of the left-hand side of (\ref{eq:phase_condition}) can be
reduced to one (e.g., $\theta_{s,2}$ when $\theta_{s,1}/\theta_{s,2}=2$), but at
the the right-hand side, both $\epsilon$ and $\theta_{m}$ are unknown at the
slave and parameters to be estimated. In other words, for the equations (5) and
(6) of \cite{chin09:_ieee} to be valid, not only the value of $\theta_{s,2}$ (or
$\theta_{s,1}$) but also the values of $\epsilon$ and $\theta_{m}$ should be
known at the slave, which verifies that Chin and Chen's formulation of
multi-parameter estimation is invalid.

Note that in case of $\theta_{s,1}=2\theta_{s,2}$ (i.e., both the slave clocks
begin (or reset) at the same time), the times of slave clocks should meet the
following condition (again, ignoring random noise components for
simplicity)\footnote{It can be generalized with a known offset between the two
  slave clocks, which is different from that with the master clock (i.e.,
  $\theta$).}:
\begin{equation}
  \label{eq:slave_clocks}
  T_{s1}(T_{m}) = 2 T_{s2}(T_{m}) .
\end{equation}
In such a case the equations (5) and (6) of \cite{chin09:_ieee} should be
rewritten as
\begingroup
\setlength{\arraycolsep}{0.0em}
\begin{eqnarray}
  T_{s1}.i &{=}& 2 \left\{\left(1 + \epsilon\right) \left(T_{m}.i + D_{m2s}\right) + \theta\right\} + \phi_{1,i} \label{eq:new_slave_clock1} \\
  T_{s2}.i &{=}& \left( 1 + \epsilon\right) \left(T_{m}.i + D_{m2s}\right) + \theta + \phi_{2,i} \label{eq:new_slave_clock2} 
\end{eqnarray}
\endgroup
Unlike the original equations of \cite{chin09:_ieee},
(\ref{eq:new_slave_clock1}) and (\ref{eq:new_slave_clock2}) cannot be
manipulated to separate the delay ($D_{m2s}$) from the clock offset ($\theta$)
or vice versa, which invalidates the resulting maximum likelihood (ML) estimates
in \cite{chin09:_ieee}.

Note that, in practical applications like clock synchronization in wireless
sensor networks (WSNs), the best one can do with these equations is the
estimation of the clock skew and the clock offset assuming that $D_{m2s}$ is
negligible and that $\epsilon \approx 0$ as suggested in
\cite{wu11:_clock_synch_wirel_sensor_networ}.

\end{document}